\documentstyle[psfig]{mn}
\newif\ifAMStwofonts
\AMStwofontstrue

\newcommand{\Msun}{\mbox{$M_{\odot}$}}
\newcommand{\sub}[1]{\mbox{$_{\rm #1}$}}
\newcommand{\Mi}{\mbox{$m\sub{i}$}}

\newcommand{\diff}{\mbox{d}}
\title[Red Clump $K$-band brightness]
{Population effects on the red giant clump
absolute magnitude: The $K$-band.}

\author[M.~Salaris \& L.~Girardi]
       {Maurizio Salaris$^{1}$ and L\'eo Girardi$^{2}$ \\
$^1$Astrophysics Research Institute, Liverpool John Moores
        University, Twelve Quays House, Egerton Wharf, Birkenhead CH41 1LD, 
	UK \\
$^2$Osservatorio Astronomico di Trieste,
	Via Tiepolo 11, I-34131 Trieste, Italy} 
\date{Accepted 2002 ???.
      Received 2002, June ??;
      in original form 2002 ???}

\pagerange{\pageref{firstpage}--\pageref{lastpage}}
\pubyear{2002}

\begin{document}

\maketitle

\label{firstpage}

%%%%%%%%%%%%%%%%%%%%%%%%%%%%%%%%%%%%%%%%%%%%%%%%
\begin{abstract}

We present a detailed analysis of the behaviour of the Red Clump
$K$-band absolute magnitude ($M_K^{\rm RC}$) 
in simple and composite stellar populations,
in light of its use as standard candle for distance determinations.
The advantage of using $M_K^{\rm RC}$, 
following recent empirical calibrations of its value
for the solar neighbourhood, arises from its very
low sensitivity to the extinction by interstellar dust. 
We show that, as in the case of the $V$- and $I$-band results,
$M_K^{\rm RC}$ is a complicated
function of the stellar metallicity $Z$ and age $t$. In general, 
$M_K^{\rm RC}$ is more sensitive to $t$ and $Z$ 
than $M_I^{\rm RC}$, for high $t$ and low $Z$.
Morever, for ages above $\sim$1.5 Gyr,  
$M_K^{\rm RC}$ decreases with increasing $Z$, the opposite
behaviour with respect to $M_V^{\rm RC}$ and $M_I^{\rm RC}$.

We provide data and equations which allow the determination of the
$K$-band population correction $\Delta M_K^{\rm RC}$
(difference between the 
Red Clump brightness in the solar neighbourhood and in the population
under scrutiny) for any generic stellar
population. These data complement the results presented in 
Girardi \& Salaris~(2001) for the $V$- and $I$-band.
We show how data from galactic open clusters consistently support 
our predicted $\Delta M_V^{\rm RC}$, $\Delta M_I^{\rm RC}$ 
and $\Delta M_K^{\rm RC}$ values.

Multiband $VIK$ population corrections for various galaxy systems 
are provided. They can be used 
in conjunction with the method devised by Alves et al.~(2002), in
order to derive simultaneously reddening and distance from the use of
$VIK$ observations of Red Clump stars. We have positively tested 
this technique on the Galactic globular cluster 47~Tuc, for which both an
empirical parallax-based main sequence fitting distance and reddening
estimates exist. We have also studied the case of using only $V$ and
$I$ photometry, recovering consistent results for both reddening and distance.
Application of this method to an OGLE-II field, and the results by
Alves et al.~(2002), confirm a LMC distance modulus of about 18.50, in
agreement with the $HST$ extragalactic distance scale zero-point.

\end{abstract}


%%%%%%%%%%%%%%%%%%%%%%%%%%%%%%%%%%%%%%%%%%%%%%%%
\section{Introduction}
\label{sec_intro}

During the last few years a 
large body of work has been focused on the
use of helium burning red clump (RC) stars as distance indicator; this
has been mainly due to the fact that
$Hipparcos$ parallaxes allows an accurate calibration of the average
RC brightness in the solar neighbourhood
(as a comparison, $Hipparcos$ parallaxes of the closest Cepheid and
RR~Lyrae stars have much larger errors, and the calibration of their
absolute magnitude has provided conflicting results,
see e.g. Feast \& Catchpole~1997; Luri et al.~1998), 
and that the RC is easily
recognizable in the colour-magnitude-diagram (CMD) of intermediate-old
stellar populations.

The determination of the absolute magnitude of the local RC in a given
pass-band $\lambda$,
$M_{\lambda, {\rm local}}^{\rm RC}$, and the apparent magnitude 
$m_\lambda^{\rm RC}$ of the
RC in a given stellar population
is not difficult,
since in both the {\em Hipparcos} database of nearby stars,
and in CMDs covering even a small fraction of a nearby galaxy, 
one finds several hundreds of clump stars, easily 
identifiable from their CMD location. As proposed by 
Stanek \& Garnavich~(1998), a non-linear least-square fit of the function 
\begin{equation}
N(m_\lambda) = a + b m_\lambda + c m_\lambda^2 + 
	d \exp\left[-\frac{(m_\lambda^{\rm RC}-m_\lambda)^2}{2\sigma_{m_\lambda}^2}\right]
\label{eq_fit}
\end{equation}
to the histogram of stars in the clump region per magnitude bin 
provides the value of 
$m_\lambda^{\rm RC}$ and its associated standard error. 
By applying this procedure to the {\em Hipparcos} database of 
nearby stars, the RC absolute brightness in the $I$-band (Cousins)
$M_I^{\rm RC}$ has been determined with accuracy of hundredths of
magnitude (Paczy\'nski \& Stanek 1998; Stanek, Zaritsky \& Harris 1998). 

Once the mean apparent magnitude of the RC in a given photometric
band, $m_{\lambda}^{\rm RC}$, 
is measured in a nearby galaxy, its absolute distance modulus 
$\mu_0=(m-M)_0$ is easily derived by means of
\begin{equation}
\mu_0 = m_\lambda^{\rm RC} - M_{\lambda, {\rm local}}^{\rm RC} - 
	A_{m_\lambda} + \Delta M_\lambda^{\rm RC}.
\label{eq_mu}
\end{equation}
In this equation, $A_{m_\lambda}$ is the 
interstellar extinction to the 
RC population of an external galaxy, and 
$\Delta M_\lambda^{\rm RC} = M_{\lambda, {\rm local}}^{\rm RC} - 
M_{\lambda, {\rm galaxy}}^{\rm RC}$ is the population effect, i.e.\ the 
difference of the mean RC absolute magnitude between 
the local and external samples of stars.
%The parameter $\sigma_I$ gives a good indication of how sharp the 
%luminosity distribution of clump stars is, whereas $a$, $b$, $c$ 
%(also derived from the fitting procedure) are constants of less interest. 

After early claims (Paczy\'nski \& Stanek 
1998; Udalski et al.\ 1998; Stanek et al.\ 1998) based mainly on the
constancy of  $M_I^{\rm RC}$ with respect to the colour $(V-I)$,
that  the population correction ($\Delta M_I^{\rm RC}$) is negligible in the
$I$-band, Cole~(1998), Girardi 
et al.\ (1998, hereafter Paper~I) and Girardi \& Salaris~(2001,
hereafter Paper~II) have conclusively demonstrated that this is not
the case. The in-depth analysis we presented in Paper~II has shown
how theoretical stellar evolution models
are able to reproduce the morphology and properties of the 
$I$-band RC brightness in Galactic open clusters, of the local 
$Hipparcos$ RC as
well as the RC in a sample of external galaxies, 
when using current estimates of the
Star-Formation-History (SFR) and Age-Metallicity-Relation (AMR) of
their stellar populations.
The models predict a non-negligible dependence of 
$M_I^{\rm RC}$ on both metallicity and age, so that 
the population correction $\Delta M_I^{\rm RC}$ must be taken into account.
In particular,
the dependence on age is complex and non-monotonic. This implies that
accurate RC distances can be determined only if one can estimate 
SFR and AMR for the stellar population under scrutiny. If there are no
determinations of these two key parameters, errors up to $\sim$0.3 mag
in the derived distance modulus are to be accepted.

In the specific case of the Large Magellanic Cloud (LMC) distance,
that sets the zero point of the extragalactic distance scale, we 
obtained a population correction in the $I$-band
$\Delta M_I^{\rm RC}$=+0.20. Unfortunately, the evaluation of
the RC distance to the LMC is plagued by what possibly are
uncertainties in the estimate of the interstellar extinction.  
Romaniello et al.\ (2000) and Sakai, Zaritsky and Kennicutt~(2000) 
obtain, respectively, $I_0^{\rm RC}=18.12\pm0.02$ and 
$I_0^{\rm RC}=18.06\pm0.02$ from $HST$ data, after deriving the extinction to
individual stars from multiband colour fitting using theoretical
model atmospheres results;  
these values, used in conjunction with the appropriate evolutionary
correction and $M_I^{\rm RC}=-0.23\pm$0.03 for the local RC, 
provide $\mu_0=18.55\pm0.04$ and $\mu_0=18.49\pm0.04$, 
respectively. On the other hand Udalski (2000) 
has obtained $I_0^{\rm RC}=17.94\pm0.05$, using different photometric
data and determining the average stellar reddenings from the
$COBE/DIRBE$ reddening maps (Schlegel, Finkbeiner \& Davis~1998),
which gives a much shorter value, $\mu_0=18.37\pm0.06$.

To overcome uncertainties related to the extinction correction, 
Alves~(2000) and Grocholski \& Sarajedini~(2002) have recently
discussed the use of the RC in the $K$-band; the main advantage 
of working in this wavelength range is a very much
reduced sensitivity to the extinction ($A_K \sim 0.2~A_I$).
Alves~(2000) has derived $M_K^{\rm RC}=-1.61\pm$0.03 for the local RC
stars with $Hipparcos$ parallaxes and spectroscopic metallicities, and
obtained a distance to the Galactic centre ($\mu_0=14.58\pm0.11$, or
$8.24\pm0.42$ Kpc) using the observed RC
$K$-band magnitude for a sample of Baade's window RC stars with the
same mean metallicity of the local sample; the underlying assumption
is that there is no potential age-effect on the $K$-band RC level.  
Grocholski \& Sarajedini~(2002) have tested the behaviour of 
$M_K^{\rm RC}$ on a sample of single-age, single-metallicity stellar 
populations, constituted by 14 Galactic
open clusters and 2 globular clusters with independent estimates of 
reddening and distances, finding a good agreement with the behaviour
predicted by the theoretical
models by Girardi et al.~(2000), which are the same used in Paper~I
and Paper~II.
Moreover, Alves et al.~(2002) have used simultaneously 
$V$, $I$ and $K$ photometry of LMC RC stars 
to determine both the  LMC distance and the
extinction for the observed RC population. 
They have demonstrated the need to apply population corrections
to the RC absolute brightness determined from the local clump,
otherwise a negative extinction is obtained; their final result
$\mu_0=18.506\pm0.033$ for the LMC ($\mu_0=18.493\pm0.033\pm0.03$
random plus systematic error, after correcting for the 
location of the observed fields with respect
to the LMC center, and including an estimate of the probable
systematic errors involved) 
and $E(B-V)=0.089\pm0.015$ was obtained making
use of evolutionary corrections derived with the techniques and models
presented in paper~II.

%An important conclusion of their work is that 
%$M_K^{\rm RC}$ is less sensitive to age and metallicity than
%$M_I^{\rm RC}$, at least within the parameter space covered by their
%sample of star clusters (ages between $\sim$ 0.6 and $\sim$12 Gyr,
%[Fe/H] ranging from $\sim-$1.15 to $\sim$0.15 dex).

In light of these results we consider 
important to investigate theoretically the 
properties of the RC in the $K$-band, as well as to assess the
simultaneous reliability of the population corrections to the $V$-, $I$-
and $K$-band for a given stellar population.

In Sect.~\ref{sec_agemet} 
we discuss in detail the 
properties of the RC in the $K$-band and its dependence on age and metallicity,
while in Sect.~\ref{sec_popcorr} we perform further tests to assess
the reliability of the population corrections predicted by theory.

In Sect.~\ref{sec_galaxies} we present multiband
population corrections for the galaxy systems discussed
in Paper~II; in addition, we will discuss the technique used by  
Alves et al.~(2002) for their LMC distance determination,
and apply it to the case of the Galactic globular cluster 47~Tuc 
(for which reddening estimates and
an accurate empirical main-sequence fitting distance do exist)
and to a field of the LMC with independent reddening determinations.

Our final conclusions are presented in Sect.~\ref{sec_conclu}. 

%%%%%%%%%%%%%%%%%%%%%%%%%%%%%%%%%%%%%%%%%%%%%%%%
\section{The theoretical red clump location in the $K$-band}
\label{sec_agemet}

As in Paper~I and Paper~II we will base the discussion of the model 
behaviour on the Girardi et al.\ (2000) set of
evolutionary tracks and isochrones. 
We remark that different models in the literature
present systematic luminosity differences for 
the core helium burning (CHeB) stars that have passed through
the helium flash -- which are the stars belonging to the observed RC
populations -- 
due mainly to the different values of the  
helium core mass at the flash (see, e.g., Castellani et al.\ 2000; 
Salaris, Cassisi \& Weiss~2002);
however, the variation of their brightness with respect to age and
metallicity is much more consistently predicted
by theory (Castellani et al.\ 2000; Salaris et al.\ 2002).
As in Paper~II, the main results of this paper will be 
based on the strictly differential use of the model predictions. 

\begin{figure}
\psfig{file=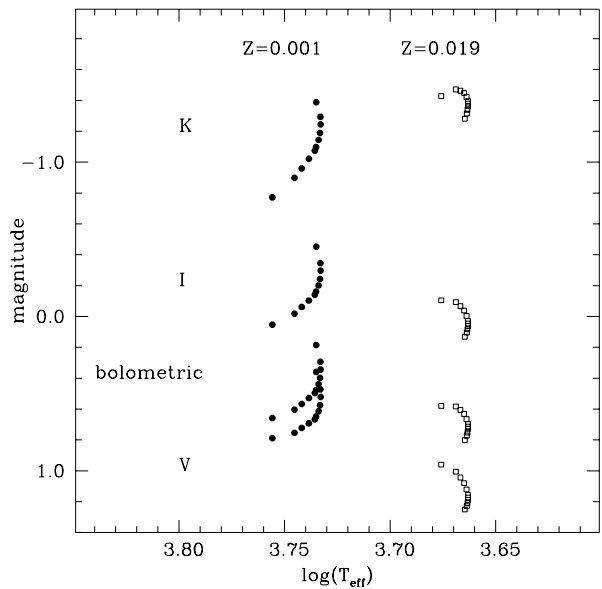,width=8.3cm}
\caption{The location of RC stars at the beginning of the
He-burning phase on an absolute magnitude--effective temperature plane,
in simple stellar populations (single-age, single-metallicity) 
of various ages (ages of 2, 3, 4, 5, 6, 8, 9, 10, 11, 13, 14 Gyr
moving clockwise; for each metallicity the oldest RC is also the 
faintest one) and two different metallicities ($Z=0.001$ and $Z=0.019$).
The age sequences are shown for, from top to bottom, $K$, $I$, 
bolometric, and $V$ magnitudes.}
\label{ZAHB}
\end{figure}

We start our analysis by showing in
Figure~\ref{ZAHB} the location of RC stars at the beginning of the
He-burning phase on a magnitude-effective temperature plane,
in simple stellar populations (single-age, single-metallicity) 
of various ages between 1 and 14 Gyr,
and two different metallicities\footnote{In the course of the paper we
will also denote the metallicity of a given stellar population with
the spectroscopic notation $[{\rm M/H}]=\log({\rm M/H})_{\rm star}-
\log({\rm M/H})_{\odot}$,
where M is the total metal abundance. [M/H] is related to $Z$ by means
of the approximate relation $[{\rm M/H}]=\log(Z/Z_{\odot})$, where we 
assume $Z_{\odot}=0.019$.
In case of a scaled-solar metal distribution $[{\rm M/H}]=[{\rm Fe/H}]$.}
($Z=0.001$ and a solar $Z=0.019$);
the behaviour of the bolometric luminosity as well as the $V$-, $I$- and
$K$-band brightness is displayed. 
One striking feature is readily apparent; while the bolometric, $V$
and $I$ brightness at fixed age increase for decreasing metallicity, 
the $K$ brightness is either approximately constant or even decreases
with decreasing metallicity.
The $V$ and $I$ magnitude follow (even if with different sensitivities)
the behaviour of the bolometric magnitude, which is essentially
determined by the variation with metallicity
of the core mass at the He-ignition; 
the higher the metallicity, the lower the core mass, the
lower the luminosity during the RC phase.
However, in case of the $K$-band the situation is completely
changed, due to the behaviour of $BC_K$.

%%%%%%%%%%%%%% figure %%%%%%%%%
%%%%%%%%%%%%%% figure %%%%%%%%%

\begin{figure}
\psfig{file=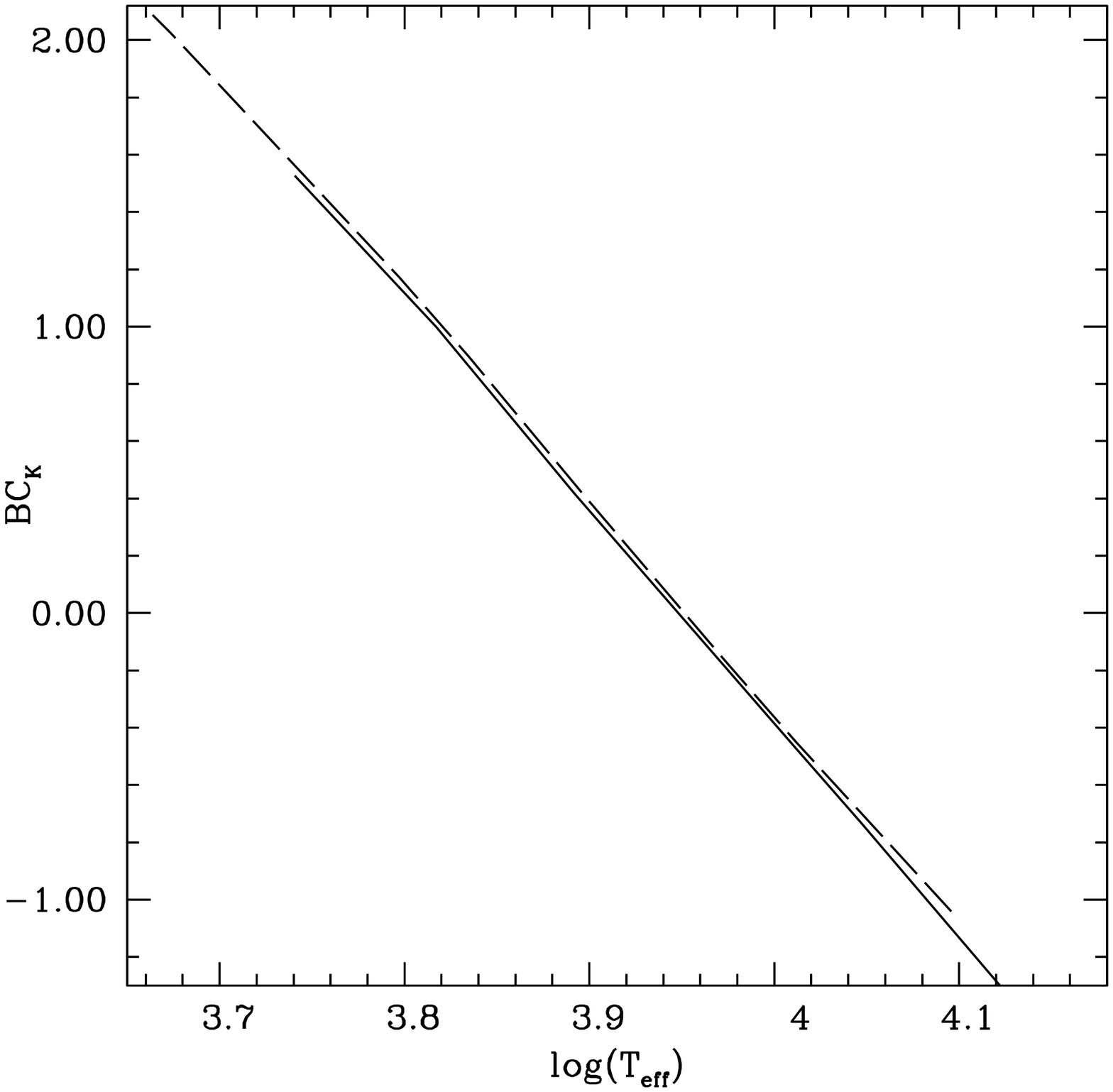,width=8.3cm}
\caption{Bolometric corrections for giants in the $K-$band,
for both $Z=0.019$ (dashed line), and $Z=0.0004$ (solid line).}
\label{BCk}
\end{figure}

Figure~\ref{BCk} displays the trend of $BC_K$
with respect to metallicity and effective temperature, for the RC stars.
It turns out that $BC_K$ is almost unaffected by Z, but it is strongly
sensitive to the value of $T_{\rm eff}$. 
Lower effective temperatures correspond 
to higher values of $BC_K$; since $M_K=M_{\rm bol}-BC_K$, cooler RC stars 
tend to be brighter. This sensitivity of $BC_K$ to $T_{\rm eff}$
is strong enough to reverse the dependence of the stellar brightness on the
He-core mass. As for the dependence of the $K$ brightness on age,
one can deduce from Figure~\ref{ZAHB}
a large effect for high ages and a reduced
one for lower ages, especially at high metallicities. This is 
a straightforward consequence of the $T_{\rm eff}$ location of the models
as a function of their age; in fact, for increasing metallicity and
decreasing age the scaling with respect to $T_{\rm eff}$ 
is reduced, thus producing a smaller variation of $BC_K$. 

It is important to remark that
the behaviour of $BC_K$ shown in Figure~\ref{BCk}
is not peculiar of the particular set of 
bolometric corrections adopted in our models (based on 
Kurucz 1993 models), but it is also found,
e.g., in the empirical results by Montegriffo et al.~(1998).

%%%%%%%%%%%%%% figure %%%%%%%%%
%%%%%%%%%%%%%% figure %%%%%%%%%
%\section{RC population corrections in the $K$-band}
%\label{sec_popcorr}

In order to explain the RC behaviour in composite stellar populations,
like the solar neighbourhood and the LMC,
we have to consider the mean RC brightness in simple stellar 
populations. It is obvious 
that the expected differential properties of the mean RC brightness
will somehow mirror the results shown in Figure~\ref{ZAHB}. 

In Paper~II we have discussed in detail how to derive the mean
RC level for a given stellar population. The most detailed approach
comes from a population synthesis algorithm; a synthetic CMD is
produced for a given stellar population model, and then 
$M_\lambda^{\rm RC}$ is derived by fitting
eq.~\ref{eq_fit} to the synthetic RC data. This is the method we 
have followed
to produce the population corrections presented in Paper~II and the
ones we discuss also in this paper.
However, it is very instructive to summarize
a simpler approach presented in Paper~II, since it
shows clearly how to decompose a complex stellar population into its
elementary 'simple' constituents.

For a given isochrone
of age and metallicity $(t,Z)$, one can perform the following integral
over the isochrone section corresponding only to CHeB stars:
\begin{equation}
\langle M_\lambda(t,Z) \rangle = -2.5\log \left[ 
        \frac{1}{N_{\rm cl}(t, Z)}
        \int^{\rm CHeB} \!\!\!\! \phi(\Mi)\, 10^{-0.4 M_\lambda}
        \diff\Mi 
	\right] 
\label{eq_inte}
\end{equation}
where $M_\lambda$ is the absolute magnitude in the pass-band 
$\lambda$, \Mi\ is the initial mass of the star at each isochrone 
point, and $\phi(\Mi)$ is the Salpeter IMF (number of stars with 
initial mass in the interval $[\Mi,\Mi+\diff\Mi]$). $N_{\rm cl}$ is the 
number of clump stars (at age $t$) per unit mass of stars 
initially born. It is simply given by the integral
of the IMF by number, along the CHeB isochrone section, i.e.\
	\begin{equation}
N_{\rm cl}(t, Z) = \int^{\rm CHeB} \phi(\Mi)\, \diff\Mi \;.
	\label{eq_ncl}
	\end{equation}
%In our case, the IMF is normalised such as to produce a 
%single-burst stellar population of total initial
%mass of 1~\Msun\ (i.e.\ $\int \Mi\,\phi(\Mi)\,\diff\Mi=1$~\Msun), 
%and a mass-to-light ratio of $M/L_V=0.2$ at an age 
%of $10^8$~yr. The details of this normalisation 
%can be found in Girardi \& Bica (1993) and 
%Salasnich et al.\ (2000). 
% It is worth remarking that the choice of the normalization
% does not affect any of the 
% results presented in this paper.
%the particular choice of IMF normalisation. 
%However, having an IMF normalised to unit mass turns out to be 
%a convenient choice.

%%%%%%%%%%%%%% figure %%%%%%%%%
\begin{figure}
\psfig{file=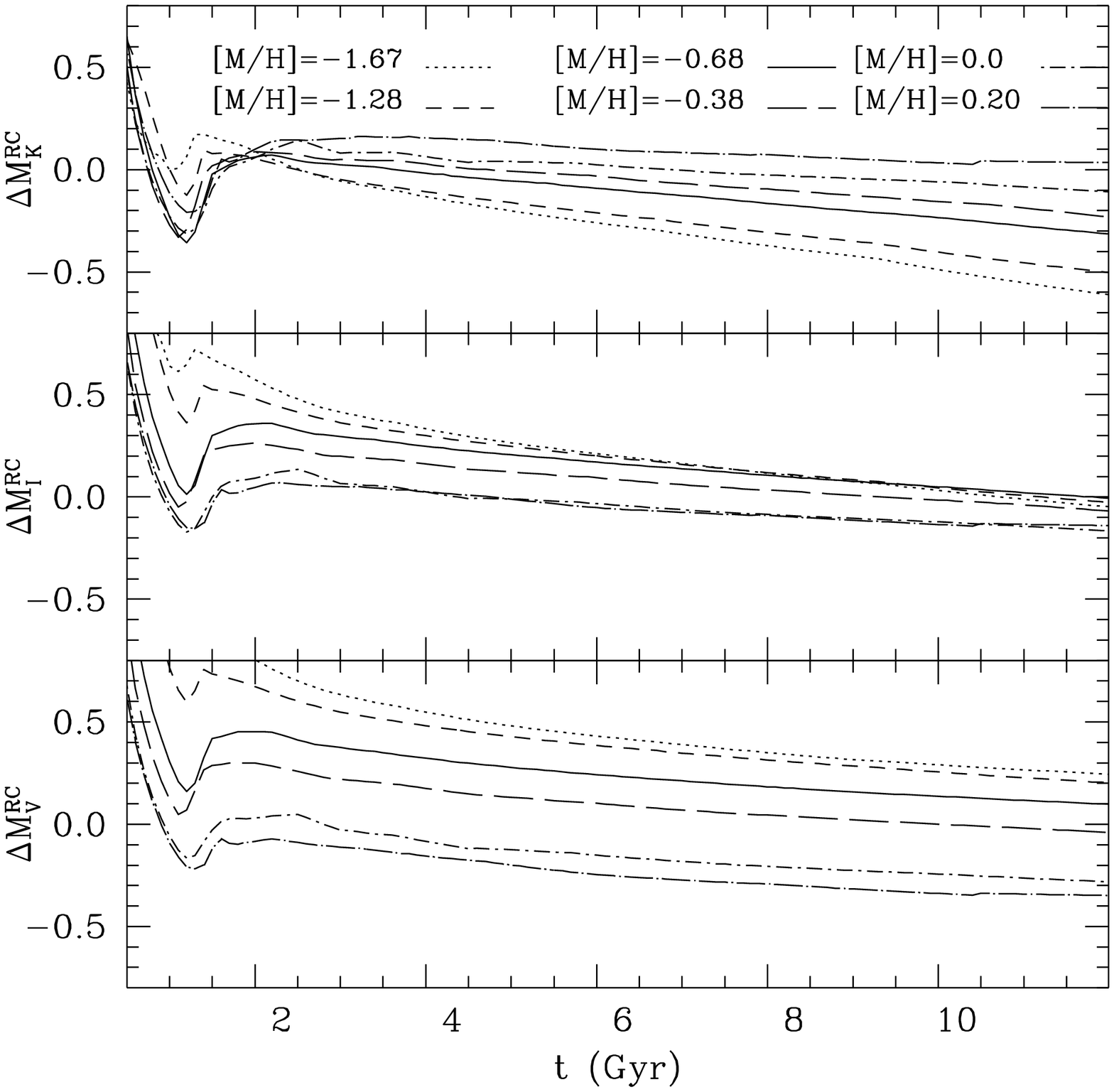,width=8.3cm}
\caption{The behaviour of the clump as a function of age, 
for several metallicities between $[{\rm M/H}]=-1.67$ and $0.20$,
($Z$ between 0.0004 and 0.03).}
\label{figmeanRC}
\end{figure} 
%%%%%%%%%%%%%% figure %%%%%%%%%

%%%%%%%%%%%%%% table %%%%%%%%%%%%%%%%%
\begin{table*}
\caption{$N_{\rm cl}$ and $\langle M_K\rangle$,
as a function of age and metallicity, from 
Girardi et al.\ (2000) isochrones.}
%\caption{Mean $K$-band absolute magnitude of clump stars, 
%$\langle M_K\rangle$,
%as a function of age and metallicity, from 
%Girardi et al.\ (2000) isochrones.}
\label{tab_mk}
\begin{tabular}{r rr rr rr rr rr rr}
\noalign{\smallskip}\hline\noalign{\smallskip}
  & \multicolumn{2}{c}{$Z=0.0004$} & \multicolumn{2}{c}{$Z=0.001$} & \multicolumn{2}{c}{$Z=0.004$}
  & \multicolumn{2}{c}{$Z=0.008$} & \multicolumn{2}{c}{$Z=0.019$} & \multicolumn{2}{c}{$Z=0.03$} \\
\noalign{\smallskip}\cline{2-3} \cline{4-5} \cline{6-7} \cline{8-9} \cline{10-11} \cline{12-13} \noalign{\smallskip}
$t$ & 
$N_{\rm cl}$ & $\langle M_K \rangle$ &
$N_{\rm cl}$ & $\langle M_K \rangle$ &
$N_{\rm cl}$ & $\langle M_K \rangle$ &
$N_{\rm cl}$ & $\langle M_K \rangle$ &
$N_{\rm cl}$ & $\langle M_K \rangle$ &
$N_{\rm cl}$ & $\langle M_K \rangle$
\\
(Gyr) & 
$(10^{-4})$ & & 
$(10^{-4})$ & & 
$(10^{-4})$ & & 
$(10^{-4})$ & & 
$(10^{-4})$ & & 
$(10^{-4})$ & 
\\
\noalign{\smallskip}\hline\noalign{\smallskip}
0.5  &  $26.30$ & $-1.969$ &  $25.70$ & $-2.175$ &  $31.60$ & $-2.191$ &	$34.30$ & $-2.047$ &  $30.60$ & $-2.024$ &  $29.40$ & $-2.118$\\
0.6  &  $29.40$ & $-1.775$ &  $30.60$ & $-2.072$ &  $37.00$ & $-1.909$ &	$40.10$ & $-1.770$ &  $36.10$ & $-1.785$ &  $33.90$ & $-1.900$\\
0.7  &  $32.30$ & $-1.686$ &  $33.70$ & $-1.941$ &  $41.40$ & $-1.693$ &	$44.90$ & $-1.599$ &  $40.20$ & $-1.607$ &  $37.30$ & $-1.733$\\
0.8  &  $36.80$ & $-1.626$ &  $37.60$ & $-1.802$ &  $44.30$ & $-1.536$ &	$48.50$ & $-1.460$ &  $43.90$ & $-1.485$ &  $40.20$ & $-1.610$\\
0.9  &  $38.00$ & $-1.601$ &  $40.90$ & $-1.665$ &  $47.20$ & $-1.417$ &	$51.80$ & $-1.350$ &  $46.50$ & $-1.388$ &  $42.90$ & $-1.505$\\
1.0  &  $43.50$ & $-1.540$ &  $44.40$ & $-1.543$ &  $50.10$ & $-1.309$ &	$54.80$ & $-1.270$ &  $48.30$ & $-1.312$ &  $45.40$ & $-1.422$\\
1.1  &  $44.30$ & $-1.546$ &  $47.00$ & $-1.456$ &  $52.80$ & $-1.222$ &	$56.70$ & $-1.209$ &  $49.90$ & $-1.254$ &  $46.60$ & $-1.364$\\
1.2  &  $41.50$ & $-1.603$ &  $48.00$ & $-1.415$ &  $53.10$ & $-1.183$ &	$54.30$ & $-1.247$ &  $49.40$ & $-1.225$ &  $47.20$ & $-1.328$\\
1.3  &  $31.00$ & $-1.711$ &  $44.00$ & $-1.478$ &  $52.20$ & $-1.236$ &	$53.50$ & $-1.363$ &  $48.40$ & $-1.251$ &  $47.10$ & $-1.333$\\
1.4  &  $20.70$ & $-1.711$ &  $28.00$ & $-1.634$ &  $51.60$ & $-1.382$ &	$27.80$ & $-1.516$ &  $48.20$ & $-1.338$ &  $46.40$ & $-1.370$\\
1.5  &  $19.40$ & $-1.699$ &  $23.00$ & $-1.619$ &  $29.60$ & $-1.517$ &	$23.20$ & $-1.559$ &  $28.20$ & $-1.446$ &  $45.80$ & $-1.496$\\
1.6  &  $18.70$ & $-1.685$ &  $21.40$ & $-1.620$ &  $22.50$ & $-1.541$ &	$20.60$ & $-1.576$ &  $25.10$ & $-1.519$ &  $31.60$ & $-1.590$\\
1.7  &  $17.80$ & $-1.675$ &  $20.10$ & $-1.616$ &  $19.70$ & $-1.563$ &	$18.80$ & $-1.596$ &  $22.20$ & $-1.554$ &  $24.30$ & $-1.576$\\
1.8  &  $16.90$ & $-1.664$ &  $19.00$ & $-1.612$ &  $18.20$ & $-1.585$ &	$17.10$ & $-1.606$ &  $19.50$ & $-1.568$ &  $20.90$ & $-1.588$\\
1.9  &  $16.00$ & $-1.650$ &  $18.00$ & $-1.607$ &  $16.60$ & $-1.595$ &	$16.30$ & $-1.617$ &  $17.40$ & $-1.580$ &  $18.90$ & $-1.615$\\
2.0  &  $15.50$ & $-1.629$ &  $16.90$ & $-1.595$ &  $16.20$ & $-1.602$ &	$15.30$ & $-1.627$ &  $16.80$ & $-1.599$ &  $17.40$ & $-1.638$\\
%2.1  &  $15.20$ & $-1.611$ &  $16.60$ & $-1.583$ &  $15.90$ & $-1.609$ &	$14.50$ & $-1.626$ &  $16.10$ & $-1.616$ &  $16.80$ & $-1.657$\\
2.2  &  $14.80$ & $-1.593$ &  $16.20$ & $-1.572$ &  $15.50$ & $-1.612$ &	$14.10$ & $-1.624$ &  $15.50$ & $-1.640$ &  $16.00$ & $-1.675$\\
%2.3  &  $13.90$ & $-1.578$ &  $15.80$ & $-1.562$ &  $15.10$ & $-1.604$ &	$13.60$ & $-1.622$ &  $15.10$ & $-1.654$ &  $15.20$ & $-1.682$\\
2.4  &  $13.50$ & $-1.559$ &  $14.90$ & $-1.550$ &  $14.80$ & $-1.595$ &	$13.30$ & $-1.619$ &  $14.70$ & $-1.668$ &  $14.40$ & $-1.682$\\
%2.5  &  $13.20$ & $-1.544$ &  $14.60$ & $-1.539$ &  $14.50$ & $-1.586$ &	$13.00$ & $-1.614$ &  $14.10$ & $-1.682$ &  $13.70$ & $-1.682$\\
2.6  &  $13.00$ & $-1.530$ &  $14.20$ & $-1.528$ &  $14.10$ & $-1.578$ &	$12.80$ & $-1.606$ &  $13.50$ & $-1.673$ &  $12.90$ & $-1.681$\\
%2.7  &  $12.80$ & $-1.516$ &  $13.90$ & $-1.518$ &  $13.50$ & $-1.572$ &	$12.50$ & $-1.598$ &  $12.90$ & $-1.659$ &  $12.30$ & $-1.683$\\
2.8  &  $12.50$ & $-1.503$ &  $13.60$ & $-1.508$ &  $13.30$ & $-1.570$ &	$12.30$ & $-1.591$ &  $12.30$ & $-1.646$ &  $12.00$ & $-1.685$\\
%2.9  &  $11.80$ & $-1.493$ &  $13.00$ & $-1.501$ &  $13.10$ & $-1.567$ &	$11.60$ & $-1.587$ &  $11.60$ & $-1.633$ &  $11.70$ & $-1.688$\\
3.0  &  $11.50$ & $-1.484$ &  $12.20$ & $-1.491$ &  $12.90$ & $-1.564$ &	$11.00$ & $-1.590$ &  $11.00$ & $-1.621$ &  $11.50$ & $-1.690$\\
%3.1  &  $11.30$ & $-1.476$ &  $12.10$ & $-1.484$ &  $12.70$ & $-1.561$ &	$10.90$ & $-1.589$ &  $10.80$ & $-1.623$ &  $11.20$ & $-1.691$\\
3.2  &  $11.10$ & $-1.468$ &  $11.90$ & $-1.478$ &  $12.50$ & $-1.558$ &	$10.70$ & $-1.588$ &  $10.80$ & $-1.623$ &  $11.00$ & $-1.700$\\
%3.3  &  $10.90$ & $-1.461$ &  $11.70$ & $-1.472$ &  $12.10$ & $-1.556$ &	$10.60$ & $-1.586$ &  $10.70$ & $-1.623$ &  $10.90$ & $-1.699$\\
3.4  &  $10.70$ & $-1.454$ &  $11.50$ & $-1.466$ &  $11.80$ & $-1.553$ &	$10.50$ & $-1.585$ &  $10.70$ & $-1.623$ &  $10.80$ & $-1.698$\\
%3.5  &  $10.50$ & $-1.447$ &  $11.30$ & $-1.461$ &  $11.70$ & $-1.548$ &	$10.40$ & $-1.584$ &  $10.60$ & $-1.623$ &  $10.70$ & $-1.697$\\
3.6  &  $10.30$ & $-1.440$ &  $11.10$ & $-1.455$ &  $11.50$ & $-1.542$ &	$10.30$ & $-1.584$ &  $10.20$ & $-1.626$ &  $10.60$ & $-1.696$\\
%3.7  &  $9.87$ & $-1.434$ &   $10.90$ & $-1.450$ &  $11.30$ & $-1.537$ &	$10.20$ & $-1.582$ &  $9.76$ & $-1.621$ &   $10.50$ & $-1.696$\\
3.8  &  $9.65$ & $-1.425$ &   $10.50$ & $-1.444$ &  $11.10$ & $-1.532$ &	$10.00$ & $-1.577$ &  $9.54$ & $-1.615$ &   $9.99$ & $-1.699$\\
%3.9  &  $9.52$ & $-1.416$ &   $10.40$ & $-1.438$ &  $11.00$ & $-1.527$ &	$9.84$ & $-1.572$ &   $9.31$ & $-1.609$ &   $9.76$ & $-1.694$\\
4.0  &  $9.38$ & $-1.408$ &   $10.30$ & $-1.432$ &  $10.80$ & $-1.523$ &	$9.67$ & $-1.567$ &   $9.09$ & $-1.603$ &   $9.63$ & $-1.693$\\
%4.1  &  $9.25$ & $-1.400$ &   $10.20$ & $-1.426$ &  $10.60$ & $-1.518$ &	$9.49$ & $-1.562$ &   $8.86$ & $-1.597$ &   $9.50$ & $-1.691$\\
%4.2  &  $9.11$ & $-1.393$ &   $10.10$ & $-1.420$ &  $9.89$ & $-1.517$ &	$9.32$ & $-1.557$ &   $8.64$ & $-1.591$ &   $9.36$ & $-1.689$\\
4.3  &  $8.98$ & $-1.385$ &   $9.93$ & $-1.414$ &	  $9.55$ & $-1.509$ &	$9.14$ & $-1.553$ &   $8.42$ & $-1.586$ &   $9.23$ & $-1.688$\\
%4.4  &  $8.84$ & $-1.378$ &   $9.81$ & $-1.409$ &	  $9.44$ & $-1.505$ &	$8.96$ & $-1.549$ &   $8.19$ & $-1.581$ &   $9.10$ & $-1.686$\\
%4.5  &  $8.71$ & $-1.371$ &   $9.69$ & $-1.403$ &	  $9.33$ & $-1.502$ &	$8.79$ & $-1.544$ &   $7.99$ & $-1.575$ &   $8.96$ & $-1.685$\\
4.6  &  $8.58$ & $-1.364$ &   $9.57$ & $-1.398$ &	  $9.21$ & $-1.498$ &	$8.29$ & $-1.541$ &   $7.88$ & $-1.581$ &   $8.83$ & $-1.683$\\
%4.7  &  $8.45$ & $-1.358$ &   $9.45$ & $-1.393$ &	  $9.10$ & $-1.495$ &	$8.10$ & $-1.541$ &   $7.83$ & $-1.581$ &   $8.46$ & $-1.684$\\
%4.8  &  $8.32$ & $-1.351$ &   $9.33$ & $-1.388$ &	  $8.99$ & $-1.492$ &	$8.01$ & $-1.538$ &   $7.77$ & $-1.581$ &   $8.20$ & $-1.680$\\
4.9  &  $8.04$ & $-1.345$ &   $9.17$ & $-1.384$ &	  $8.87$ & $-1.488$ &	$7.93$ & $-1.535$ &   $7.72$ & $-1.581$ &   $8.07$ & $-1.675$\\
%5.0  &  $7.96$ & $-1.339$ &   $8.95$ & $-1.379$ &	  $8.76$ & $-1.485$ &	$7.85$ & $-1.533$ &   $7.67$ & $-1.580$ &   $7.94$ & $-1.671$\\
%5.1  &  $7.88$ & $-1.332$ &   $8.85$ & $-1.373$ &	  $8.65$ & $-1.482$ &	$7.76$ & $-1.530$ &   $7.61$ & $-1.579$ &   $7.82$ & $-1.668$\\
5.2  &  $7.80$ & $-1.326$ &   $8.74$ & $-1.368$ &	  $8.54$ & $-1.479$ &	$7.68$ & $-1.527$ &   $7.56$ & $-1.578$ &   $7.69$ & $-1.664$\\
%5.3  &  $7.72$ & $-1.320$ &   $8.64$ & $-1.363$ &	  $8.42$ & $-1.476$ &	$7.60$ & $-1.525$ &   $7.50$ & $-1.577$ &   $7.56$ & $-1.660$\\
%5.4  &  $7.64$ & $-1.314$ &   $8.54$ & $-1.358$ &	  $8.22$ & $-1.474$ &	$7.52$ & $-1.522$ &   $7.45$ & $-1.576$ &   $7.43$ & $-1.656$\\
5.5  &  $7.55$ & $-1.308$ &   $8.44$ & $-1.352$ &	  $7.72$ & $-1.469$ &	$7.43$ & $-1.520$ &   $7.39$ & $-1.574$ &   $7.31$ & $-1.653$\\
%5.6  &  $7.47$ & $-1.302$ &   $8.34$ & $-1.347$ &	  $7.64$ & $-1.464$ &	$7.35$ & $-1.517$ &   $7.34$ & $-1.572$ &   $7.18$ & $-1.649$\\
%5.7  &  $7.39$ & $-1.297$ &   $8.24$ & $-1.343$ &	  $7.57$ & $-1.460$ &	$7.27$ & $-1.515$ &   $6.82$ & $-1.574$ &   $7.05$ & $-1.646$\\
%5.8  &  $7.31$ & $-1.291$ &   $8.14$ & $-1.338$ &	  $7.50$ & $-1.456$ &	$7.18$ & $-1.512$ &   $6.53$ & $-1.570$ &   $6.93$ & $-1.642$\\
%5.9  &  $7.23$ & $-1.286$ &   $8.04$ & $-1.333$ &	  $7.43$ & $-1.453$ &	$6.88$ & $-1.512$ &   $6.46$ & $-1.567$ &   $6.80$ & $-1.639$\\
6.0  &  $7.15$ & $-1.280$ &   $7.94$ & $-1.328$ &	  $7.36$ & $-1.449$ &	$6.47$ & $-1.508$ &   $6.38$ & $-1.564$ &   $6.46$ & $-1.637$\\
%6.1  &  $7.07$ & $-1.275$ &   $7.84$ & $-1.324$ &	  $7.29$ & $-1.445$ &	$6.41$ & $-1.504$ &   $6.31$ & $-1.561$ &   $6.26$ & $-1.634$\\
%6.2  &  $6.99$ & $-1.270$ &   $7.74$ & $-1.319$ &	  $7.22$ & $-1.441$ &	$6.35$ & $-1.500$ &   $6.24$ & $-1.558$ &   $6.21$ & $-1.632$\\
%6.3  &  $6.91$ & $-1.265$ &   $7.64$ & $-1.315$ &	  $7.15$ & $-1.438$ &	$6.28$ & $-1.496$ &   $6.17$ & $-1.555$ &   $6.16$ & $-1.631$\\
%6.4  &  $6.83$ & $-1.260$ &   $7.54$ & $-1.311$ &	  $7.08$ & $-1.434$ &	$6.22$ & $-1.492$ &   $6.11$ & $-1.552$ &   $6.11$ & $-1.629$\\
6.5  &  $6.75$ & $-1.255$ &   $7.44$ & $-1.306$ &	  $7.01$ & $-1.430$ &	$6.16$ & $-1.488$ &   $6.04$ & $-1.550$ &   $6.06$ & $-1.628$\\
%6.6  &  $6.66$ & $-1.250$ &   $7.34$ & $-1.302$ &	  $6.93$ & $-1.427$ &	$6.10$ & $-1.485$ &   $5.97$ & $-1.547$ &   $6.01$ & $-1.626$\\
%6.7  &  $6.60$ & $-1.244$ &   $6.83$ & $-1.300$ &	  $6.86$ & $-1.424$ &	$6.04$ & $-1.482$ &   $5.91$ & $-1.544$ &   $5.96$ & $-1.625$\\
%6.8  &  $6.55$ & $-1.238$ &   $6.73$ & $-1.292$ &	  $6.79$ & $-1.420$ &	$5.99$ & $-1.478$ &   $5.84$ & $-1.542$ &   $5.92$ & $-1.623$\\
%6.9  &  $6.51$ & $-1.232$ &   $6.68$ & $-1.287$ &	  $6.72$ & $-1.417$ &	$5.93$ & $-1.475$ &   $5.77$ & $-1.539$ &   $5.87$ & $-1.622$\\
7.0  &  $6.46$ & $-1.225$ &   $6.63$ & $-1.282$ &	  $6.65$ & $-1.413$ &	$5.87$ & $-1.472$ &   $5.71$ & $-1.536$ &   $5.82$ & $-1.621$\\
%7.1  &  $6.41$ & $-1.219$ &   $6.57$ & $-1.277$ &	  $6.58$ & $-1.410$ &	$5.82$ & $-1.469$ &   $5.64$ & $-1.534$ &   $5.78$ & $-1.619$\\
%7.2  &  $6.36$ & $-1.213$ &   $6.52$ & $-1.272$ &	  $6.51$ & $-1.407$ &	$5.76$ & $-1.466$ &   $5.58$ & $-1.531$ &   $5.73$ & $-1.618$\\
%7.3  &  $6.31$ & $-1.207$ &   $6.46$ & $-1.267$ &	  $6.38$ & $-1.404$ &	$5.70$ & $-1.463$ &   $5.51$ & $-1.528$ &   $5.68$ & $-1.617$\\
%7.4  &  $6.26$ & $-1.201$ &   $6.41$ & $-1.262$ &	  $6.27$ & $-1.399$ &	$5.65$ & $-1.460$ &   $5.44$ & $-1.526$ &   $5.63$ & $-1.615$\\
7.5  &  $6.22$ & $-1.196$ &   $6.36$ & $-1.257$ &	  $6.23$ & $-1.395$ &	$5.59$ & $-1.457$ &   $5.32$ & $-1.524$ &   $5.59$ & $-1.614$\\
%7.6  &  $6.17$ & $-1.190$ &   $6.30$ & $-1.252$ &	  $6.18$ & $-1.391$ &	$5.54$ & $-1.454$ &   $5.26$ & $-1.521$ &   $5.54$ & $-1.613$\\
%7.7  &  $6.12$ & $-1.184$ &   $6.25$ & $-1.247$ &	  $6.13$ & $-1.387$ &	$5.48$ & $-1.451$ &   $5.23$ & $-1.519$ &   $5.49$ & $-1.612$\\
%7.8  &  $6.07$ & $-1.179$ &   $6.19$ & $-1.242$ &	  $6.08$ & $-1.383$ &	$5.42$ & $-1.448$ &   $5.19$ & $-1.517$ &   $5.43$ & $-1.610$\\
%7.9  &  $6.02$ & $-1.173$ &   $6.14$ & $-1.238$ &	  $6.03$ & $-1.379$ &	$5.37$ & $-1.445$ &   $5.16$ & $-1.516$ &   $5.34$ & $-1.613$\\
8.0  &  $5.97$ & $-1.168$ &   $6.08$ & $-1.233$ &	  $5.99$ & $-1.375$ &	$5.32$ & $-1.446$ &   $5.12$ & $-1.514$ &   $5.31$ & $-1.611$\\
%8.1  &  $5.93$ & $-1.163$ &   $6.03$ & $-1.229$ &	  $5.94$ & $-1.371$ &	$5.29$ & $-1.442$ &   $5.09$ & $-1.512$ &   $5.28$ & $-1.608$\\
%8.2  &  $5.88$ & $-1.157$ &   $5.98$ & $-1.224$ &	  $5.89$ & $-1.368$ &	$5.26$ & $-1.439$ &   $5.05$ & $-1.510$ &   $5.24$ & $-1.606$\\
%8.3  &  $5.83$ & $-1.152$ &   $5.92$ & $-1.220$ &	  $5.85$ & $-1.364$ &	$5.23$ & $-1.436$ &   $5.01$ & $-1.508$ &   $5.21$ & $-1.604$\\
%8.4  &  $5.78$ & $-1.147$ &   $5.87$ & $-1.216$ &	  $5.80$ & $-1.360$ &	$5.20$ & $-1.433$ &   $4.98$ & $-1.507$ &   $5.18$ & $-1.602$\\
%8.5  &  $5.73$ & $-1.142$ &   $5.82$ & $-1.211$ &	  $5.75$ & $-1.357$ &	$5.17$ & $-1.429$ &   $4.94$ & $-1.505$ &   $5.14$ & $-1.600$\\
%8.6  &  $5.69$ & $-1.137$ &   $5.76$ & $-1.207$ &	  $5.71$ & $-1.353$ &	$5.14$ & $-1.426$ &   $4.91$ & $-1.503$ &   $5.11$ & $-1.598$\\
%8.7  &  $5.64$ & $-1.132$ &   $5.71$ & $-1.203$ &	  $5.66$ & $-1.349$ &	$5.11$ & $-1.423$ &   $4.87$ & $-1.502$ &   $5.08$ & $-1.596$\\
%8.8  &  $5.59$ & $-1.127$ &   $5.65$ & $-1.199$ &	  $5.61$ & $-1.346$ &	$5.08$ & $-1.420$ &   $4.84$ & $-1.500$ &   $5.05$ & $-1.594$\\
%8.9  &  $5.54$ & $-1.122$ &   $5.60$ & $-1.195$ &	  $5.56$ & $-1.343$ &	$5.05$ & $-1.417$ &   $4.80$ & $-1.498$ &   $5.01$ & $-1.592$\\
9.0  &  $5.49$ & $-1.118$ &   $5.55$ & $-1.190$ &	  $5.52$ & $-1.339$ &	$5.02$ & $-1.414$ &   $4.77$ & $-1.497$ &   $4.98$ & $-1.590$\\
%9.1  &  $5.45$ & $-1.113$ &   $5.49$ & $-1.186$ &	  $5.47$ & $-1.336$ &	$4.99$ & $-1.411$ &   $4.73$ & $-1.495$ &   $4.95$ & $-1.588$\\
%9.2  &  $5.10$ & $-1.110$ &   $5.37$ & $-1.183$ &	  $5.43$ & $-1.332$ &	$4.96$ & $-1.408$ &   $4.70$ & $-1.493$ &   $4.92$ & $-1.586$\\
%9.3  &  $4.95$ & $-1.104$ &   $5.25$ & $-1.178$ &	  $5.38$ & $-1.329$ &	$4.93$ & $-1.405$ &   $4.66$ & $-1.492$ &   $4.89$ & $-1.584$\\
%9.4  &  $4.94$ & $-1.096$ &   $5.22$ & $-1.172$ &	  $5.33$ & $-1.326$ &	$4.90$ & $-1.402$ &   $4.62$ & $-1.490$ &   $4.85$ & $-1.582$\\
%9.5  &  $4.93$ & $-1.088$ &   $5.19$ & $-1.166$ &	  $5.29$ & $-1.323$ &	$4.86$ & $-1.399$ &   $4.59$ & $-1.489$ &   $4.82$ & $-1.580$\\
%9.6  &  $4.91$ & $-1.081$ &   $5.17$ & $-1.160$ &	  $5.24$ & $-1.319$ &	$4.83$ & $-1.396$ &   $4.55$ & $-1.487$ &   $4.79$ & $-1.578$\\
%9.7  &  $4.90$ & $-1.074$ &   $5.14$ & $-1.154$ &	  $5.19$ & $-1.316$ &	$4.80$ & $-1.393$ &   $4.52$ & $-1.486$ &   $4.76$ & $-1.576$\\
%9.8  &  $4.88$ & $-1.066$ &   $5.11$ & $-1.148$ &	  $5.15$ & $-1.313$ &	$4.77$ & $-1.390$ &   $4.48$ & $-1.484$ &   $4.73$ & $-1.574$\\
%9.9  &  $4.87$ & $-1.059$ &   $5.09$ & $-1.142$ &	  $5.11$ & $-1.309$ &	$4.74$ & $-1.387$ &   $4.45$ & $-1.483$ &   $4.69$ & $-1.572$\\
10.0  &  $4.85$ & $-1.052$ &   $5.06$ & $-1.137$ &	  $5.14$ & $-1.306$ &	$4.71$ & $-1.385$ &   $4.48$ & $-1.480$ &   $4.66$ & $-1.571$\\
%10.1  &  $4.84$ & $-1.045$ &   $5.04$ & $-1.131$ &	  $5.12$ & $-1.303$ &	$4.68$ & $-1.382$ &   $4.51$ & $-1.480$ &   $4.63$ & $-1.569$\\
%10.2  &  $4.83$ & $-1.038$ &   $5.01$ & $-1.126$ &	  $5.10$ & $-1.299$ &	$4.65$ & $-1.379$ &   $4.49$ & $-1.477$ &   $4.60$ & $-1.567$\\
%10.3  &  $4.81$ & $-1.031$ &   $4.98$ & $-1.120$ &	  $5.07$ & $-1.294$ &	$4.61$ & $-1.376$ &   $4.46$ & $-1.474$ &   $4.57$ & $-1.565$\\
%10.4  &  $4.80$ & $-1.025$ &   $4.96$ & $-1.115$ &	  $5.05$ & $-1.290$ &	$4.58$ & $-1.374$ &   $4.44$ & $-1.472$ &   $4.58$ & $-1.563$\\
%10.5  &  $4.78$ & $-1.018$ &   $4.93$ & $-1.110$ &	  $5.03$ & $-1.286$ &	$4.55$ & $-1.371$ &   $4.42$ & $-1.469$ &   $4.58$ & $-1.580$\\
%10.6  &  $4.77$ & $-1.012$ &   $4.91$ & $-1.105$ &	  $5.01$ & $-1.281$ &	$4.52$ & $-1.369$ &   $4.40$ & $-1.466$ &   $4.57$ & $-1.580$\\
%10.7  &  $4.75$ & $-1.005$ &   $4.88$ & $-1.099$ &	  $4.98$ & $-1.277$ &	$4.58$ & $-1.365$ &   $4.37$ & $-1.464$ &   $4.56$ & $-1.579$\\
%10.8  &  $4.74$ & $-0.999$ &   $4.85$ & $-1.094$ &	  $4.96$ & $-1.273$ &	$4.70$ & $-1.363$ &   $4.35$ & $-1.461$ &   $4.55$ & $-1.578$\\
%10.9  &  $4.72$ & $-0.993$ &   $4.83$ & $-1.089$ &	  $4.94$ & $-1.269$ &	$4.68$ & $-1.358$ &   $4.33$ & $-1.458$ &   $4.54$ & $-1.578$\\
11.0  &  $4.71$ & $-0.986$ &   $4.80$ & $-1.084$ &	  $4.91$ & $-1.265$ &	$4.65$ & $-1.354$ &   $4.31$ & $-1.456$ &   $4.53$ & $-1.577$\\
%11.1  &  $4.69$ & $-0.980$ &   $4.78$ & $-1.079$ &	  $4.89$ & $-1.260$ &	$4.63$ & $-1.349$ &   $4.29$ & $-1.453$ &   $4.52$ & $-1.576$\\
%11.2  &  $4.68$ & $-0.974$ &   $4.75$ & $-1.074$ &	  $4.87$ & $-1.256$ &	$4.61$ & $-1.344$ &   $4.26$ & $-1.451$ &   $4.51$ & $-1.576$\\
%11.3  &  $4.66$ & $-0.968$ &   $4.73$ & $-1.069$ &	  $4.85$ & $-1.252$ &	$4.58$ & $-1.340$ &   $4.24$ & $-1.448$ &   $4.50$ & $-1.575$\\
%11.4  &  $4.65$ & $-0.963$ &   $4.70$ & $-1.065$ &	  $4.82$ & $-1.248$ &	$4.56$ & $-1.335$ &   $4.22$ & $-1.446$ &   $4.49$ & $-1.575$\\
%11.5  &  $4.64$ & $-0.957$ &   $4.67$ & $-1.060$ &	  $4.80$ & $-1.244$ &	$4.54$ & $-1.331$ &   $4.20$ & $-1.443$ &   $4.47$ & $-1.574$\\
%11.6  &  $4.62$ & $-0.951$ &   $4.65$ & $-1.055$ &	  $4.78$ & $-1.241$ &	$4.51$ & $-1.326$ &   $4.17$ & $-1.441$ &   $4.46$ & $-1.574$\\
%11.7  &  $4.61$ & $-0.945$ &   $4.62$ & $-1.051$ &	  $4.75$ & $-1.237$ &	$4.49$ & $-1.322$ &   $4.15$ & $-1.438$ &   $4.45$ & $-1.573$\\
%11.8  &  $4.59$ & $-0.940$ &   $4.60$ & $-1.046$ &	  $4.73$ & $-1.233$ &	$4.47$ & $-1.317$ &   $4.13$ & $-1.436$ &   $4.44$ & $-1.573$\\
%11.9  &  $4.58$ & $-0.934$ &   $4.57$ & $-1.041$ &	  $4.71$ & $-1.229$ &	$4.44$ & $-1.313$ &   $4.11$ & $-1.434$ &   $4.43$ & $-1.572$\\
12.0  &  $4.56$ & $-0.929$ &   $4.55$ & $-1.037$ &	  $4.69$ & $-1.225$ &	$4.42$ & $-1.309$ &   $4.09$ & $-1.431$ &   $4.42$ & $-1.572$\\
\noalign{\smallskip}\hline\noalign{\smallskip}
\end{tabular}

\end{table*}
%%%%%%%%%%%%%% 

{From} eq.~\ref{eq_inte}, accurate values for the 
mean clump magnitudes can be determined, 
in case of a simple stellar population.
These mean magnitudes enter straightforwardly
in the computation of 
%(ii) The intrinsic dispersion of clump magnitudes is 
%slightly larger for younger ages and lower metallicities. 
%For high metallicities and old ages, the LF 
%of clump stars is characterized by a sharp spike that 
%coincides with the ZAHB bin,
%and a decreasing tail for higher luminosities. Thus, the mean
%values $\langle M_\lambda\rangle$ are always brighter than the 
%maximum of the LF. This offset is of order 0.1~mag, for the 
%several ages and metallicities considered in the figure.
%
%Therefore, reducing the clump of a single generation of stars
%to a single point of magnitude $\langle M_\lambda(t,Z)\rangle$,
%should be seen as a useful approximation, rather than the
%detailed behaviour indicated by models.
the mean clump 
magnitude for a given galaxy model, of total age $T$. In fact,  
in this case one needs to perform the following integral:
	\begin{equation}
\langle M_\lambda({\rm gal}) \rangle = 
	\frac{1}{N_{\rm cl}({\rm gal})}
	\int_{t=0}^T N_{\rm cl}(t,Z)\, \psi(t)\, \langle M_\lambda(t,Z)\rangle
	\,\diff t \;,
	\label{eq_integal}
	\end{equation}
where
	\begin{equation}
N_{\rm cl}({\rm gal}) = 
	\int_{t=0}^T N_{\rm cl}(t, Z)\, \psi(t)\, \diff t \;.
	\label{eq_nclgal}
	\end{equation}

The function $\psi(t)$ is the SFR
(in \Msun\ by unit time) at a moment $t$ in the past,
for the galaxy model considered; in addition,
also the AMR $Z(t)$ should
be specified. One can average the magnitudes in 
eq.~\ref{eq_integal}, instead of luminosities as in the previous
eq.~\ref{eq_inte}; the reason is, as explained in Paper~II, 
that in this way
one obtains a quantity similar to the $M_\lambda^{\rm RC}$ derived 
by means of eq.~\ref{eq_fit}.
As discussed in Paper~II, the population corrections evaluated with
this approach are in very good agreement with the results obtained from the
population synthesis approach; they are directly evaluated from the
average RC brightness of the elementary simple populations,
whose properties are highlighted in the following.

Table~\ref{tab_mk} presents the $K$-band mean absolute magnitude 
of clump stars, $\langle M_K\rangle$,
as a function of age and metallicity, as derived from 
Girardi et al.\ (2000) isochrones by means of
eq.~\ref{eq_inte}, together with the values of $N_{\rm cl}(t, Z)$.
%provided in Table~1 of Paper~II,
This table can be used to easily derive the population 
correction in the $K$-band for any galaxy model.
The same information is provided for the $V$- and $I$-band
in Table~1 of Paper II. It is important to remark the pronounced
maximum of $N_{\rm cl}$ for ages between 1 and 2 Gyr, which causes
these relatively younger clump stars to be very numerous in 
galaxies with recent star formation. This aspect is thoroughly 
discussed in Paper II, and in Girardi (1999).

Figure~\ref{figmeanRC} portrays the behaviour of the mean RC magnitude 
in simple stellar populations of different metallicities and ages, for
the $V$, $I$ and $K$ passbands. In line with our desire to use the models
only in a differential way, we have actually plotted the theoretical
population correction (not the actual mean RC magnitude) 
for any given simple population considered.
The theoretical counterpart of the local RC has been computed using
the SFR and AMR by Rocha-Pinto et al.~(2000a,b -- see Paper~II for
further discussions on this issue). 

It is evident that the behaviour
of the population corrections -- and therefore of the mean RC brightness
-- is strongly dependent on the photometric band used.
For ages larger than about 1.5 Gyr 
$\Delta M_K^{\rm RC}$ increases for
increasing metallicity (hence $M_K^{\rm RC}$ decreases), 
which is exactly the contrary of 
what happens to $\Delta M_V^{\rm RC}$ and $\Delta M_I^{\rm RC}$.
If we consider the entire [M/H] range spanned by our models 
(from $-$1.7 up to 0.20),
for ages below $\sim$4 Gyr $M_K^{\rm RC}$ is on average 
less sensitive to [M/H] than both $M_V^{\rm RC}$ and $M_I^{\rm RC}$. 
However, when the age is larger than this limit
$M_K^{\rm RC}$ is
more affected by [M/H] than $M_I^{\rm RC}$; for ages larger than
$\sim$11 Gyr, $M_K^{\rm RC}$
is more metallicity-dependent than also $M_V^{\rm RC}$, 
at least in case of [M/H] lower than $\sim -$0.7.
In the [M/H] interval between 0.0 and
$\sim -$0.4 (this lower limit corresponds approximately to the average
metallicity of the LMC clump stars) $M_K^{\rm RC}$ is less affected 
by the metallicity than 
both $M_I^{\rm RC}$ and $M_V^{\rm RC}$, at least up to ages of about
11 Gyr. At super-solar [M/H] and for ages above $\sim$3 Gyr,  
$M_K^{\rm RC}$ is more sensitive to metallicity than the
$V$ and $I$ brightness.

This complex relationship among the RC $V$, $I$ and $K$ brightness
stems from the different sensitivities to age. In particular,
$M_K^{\rm RC}$ is more affected by age than both $M_I^{\rm RC}$
and $M_V^{\rm RC}$, for the lower metallicities and higher ages.

\section{Testing the theoretical multicolour population corrections}
\label{sec_popcorr}

Paper~II and Grocholski \& Sarajedini~(2002) have already 
shown separately the consistency of the theoretical results 
in the $I$ and $K$ band,
when applied to a
sample of Galactic open clusters of various ages and metallicities. 
The very different behaviour of the population corrections in
different passbands allows one to perform the following
additional test for the accuracy of the models.

\begin{figure}
\psfig{file=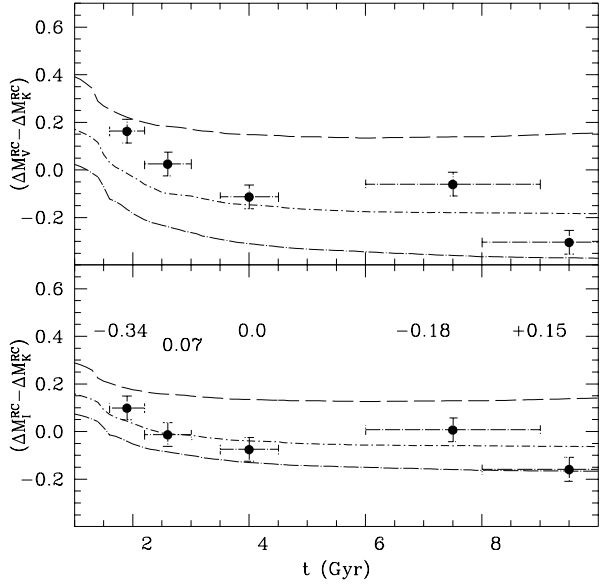,width=8.3cm}
\caption{The population correction difference between two different 
photometric bands, as predicted by theory (lines) and from observations 
of a sample of galactic open clusters (dots) with photometry in 
$V$, $I$ and $K$. The theoretical metallicities are 
$[{\rm M/H}]=-0.38$ (long-dashed line), $0$ (dot-short-dashed), and 
$0.20$ (dot-long-dashed). The numbers on the plot indicate the 
metallicity [M/H] for each cluster; the clusters are, from left to
right, NGC~2204, NGC~6819, M~67, Be~39 and NGC~6791.}
\label{clusters}
\end{figure}

Figure~\ref{clusters} compares the difference
of the population corrections in two different photometric bands, 
between theory and observations, for a sample of Galactic open clusters with
photometry in $V$, $I$ and $K$. 
The data about the clusters $V$ and $I$  
magnitudes, as well as ages, [Fe/H] and reddening values
are taken from Sarajedini~(1999). $K$-band photometric data are from 
Grocholski \& Sarajedini~(2002). The clusters' ages
have been derived with respect to the reference cluster M~67, assumed to be 
4 Gyr old on the base of previous studies; the age scaling
has been obtained from the distance modulus difference
(derived from the Main Sequence-fitting)
with respect to M~67. 
It is important to notice that the difference of the evolutionary 
corrections in two different photometric bands is independent of the
adopted distance modulus (while the cluster age is dependent on the
relative Main Sequence-fitting distance moduli and the age zero-point).
As for the empirical calibration of the local RC we have used the
values provided by Alves et al.~(2002), namely
$M_{K}^{\rm RC}=-1.60\pm0.03$,
$M_{I}^{\rm RC}=-0.26\pm0.03$, $M_{V}^{\rm RC}=0.73\pm0.03$
(the theoretical values we obtain from our simulation are
$M_{K}^{\rm RC}=-1.54$, $M_{I}^{\rm RC}=-0.17$, $M_{V}^{\rm RC}=0.84$;
we stress again that nowhere we use individual absolute magnitudes from the
stellar models, but only magnitude differences).
The value $M_{I}^{\rm RC}=-0.26\pm0.03$ agrees, within 1$\sigma$, with the
previous determination by Stanek \& Garnavich~(1998).

%%%%%%%%%%%%%% figure %%%%%%%%%
%%%%%%%%%%%%%% figure %%%%%%%%%

The agreement between theory and observations is satisfactory 
-- also considering the uncertainty of the order of $\sim$0.1 dex 
associated to the individual cluster [M/H] estimates --,
another confirmation for the population corrections predicted by the
stellar models we employ. In fact, taking into account the very
different behaviour of the predicted evolutionary
corrections in $V$, $I$ and $K$, it is highly improbable that a
combination of errors in the theory and in the adopted clusters'
parameters, were to produce such a good agreement.

Another interesting test makes use of the available $V$- and 
$K$-photometry of the few RC stars populating the
$Hipparcos$ database for the Hyades and Praesepe clusters. 
There is a large body of work suggesting that these two clusters
share the same age and metallicity (e.g., the discussion and
references in van Leeuwen~1999; Castellani et al.~2002). From the 
$Hipparcos$ data the cluster distances result to be
$\mu_0^{\rm Hyades}=3.33\pm0.01$ and  
$\mu_0^{\rm Praesepe}=6.37\pm0.15$ (Perryman et al.~1997;
van Leeuwen~1999); a comparison of their CMD corrected 
for the different distance moduli, as performed by
Castellani et al.~(2002), confirms a basically identical metallicity
and age, since the good overlap of the upper 
main sequence and turn off, and also of the RC region. 

\begin{figure}
\psfig{file=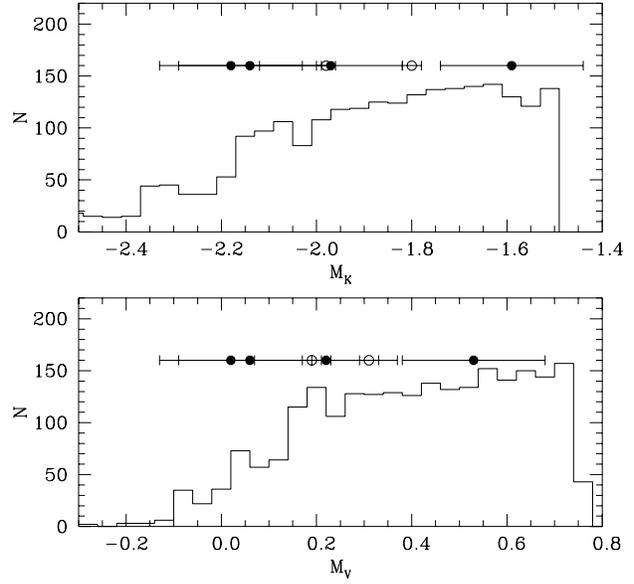,width=8.3cm}
\caption{The theoretical luminosity function
(number of stars per magnitude bin)
for Hyades RC stars in the $V$- and $K$-band
(the normalization is arbitrary); the brightness 
of the RC has the appropriate mean value
obtained after applying the theoretical population
corrections (which are of the order of $\sim-0.3$ mag in both passbands) 
to the local RC absolute magnitude.
On the same plot we show the $V$ and $K$ absolute magnitude of the 6 stars 
(2 in the Hyades and 4 in Praesepe; open and closed circles, respectively) 
populating the RC of the composite CMD of these two clusters (see text
for details).}
\label{hyades}
\end{figure}

For a common age $t=625\pm50$ million years 
(higher than the lower limit for the existence of the RC, 
which is of about 500 million years) and $[{\rm Fe/H}]=0.14$ 
(Perryman et al.~1997), we have determined the 
appropriate population correction and plotted in 
Figure~\ref{hyades} the theoretical luminosity function
(number of stars per magnitude bin)
for RC stars in the $V$- and $K$-band
(the normalization is arbitrary); the brightness 
of the RC has the appropriate mean value
obtained after applying the theoretical population
corrections (which are of the order of $\sim-0.3$ mag in both passbands) 
to the observed local RC absolute magnitude.
On the same plot we show the $V$ and $K$ absolute magnitude of the 6 stars 
(2 in the Hyades and 4 in Praesepe) populating the RC of the composite
CMD of these two clusters. Due to the small size of the star sample 
it is not possible
to determine a meaningful mean brightness to compare with theory, but 
one can notice that the empirical stellar absolute magnitudes
are in good agreement with the location and the width of
the RC as predicted by the theoretical population corrections.

As a warning, it is also important to take into account the fact 
that the values of $\Delta M_K^{\rm RC}$
and $\Delta M_V^{\rm RC}$ are both strong function of $t$ in this age
range (in case Praesepe is younger than the Hyades by $\sim$ 100 million years, its 
RC would be on average brighter by $\sim$0.2 mag than the Hyades RC);
we notice however that here is no obvious offset between the two sets 
of observational points, and between them and the theoretical
luminosity function.
In conclusion, we regard the agreement between theoretical and
empirical data in Figure~\ref{hyades} as an additional
indication of the accuracy of the 
population corrections predicted by theory.

%%%%%%%%%%%%%% figure %%%%%%%%%
%%%%%%%%%%%%%% figure %%%%%%%%%

The generally small values of $\Delta M_K^{\rm RC}$ 
for ages between $\sim$1.5 and $\sim$4 Gyr, as displayed in 
Figure~\ref{figmeanRC}, raise the question of to what extent 
one is allowed to disregard the population corrections when
considering the $K$-band RC magnitude as distance indicator
(see also the discussion in Grocholski \& Sarajedini~2002).
In this age range, the lower sensitivity of
the $K$-band RC brightness to both age and metallicity may suggest
to use the local $K$-band RC brightness as a perfect standard candle,
without taking into account the appropriate $\Delta M_K^{\rm RC}$.

%%%%%%%%%%%%%% figure %%%%%%%%%
\begin{figure}
\psfig{file=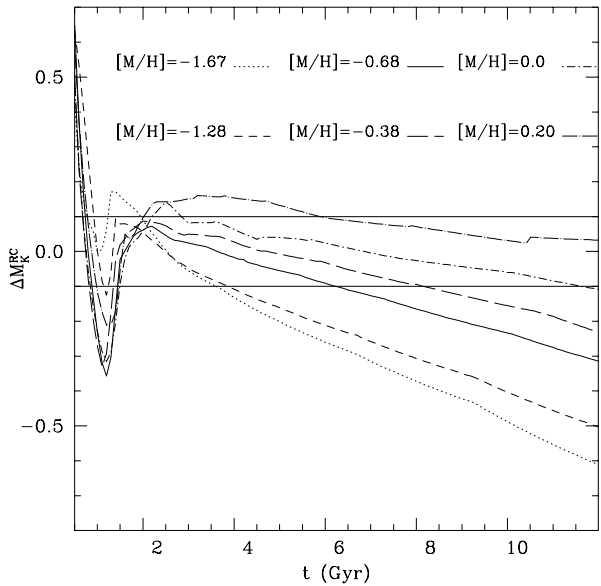,width=8.3cm}
\caption{The population corrections
to the $K$-band for simple stellar populations of varying 
age and metallicity, marking the region for which the population 
correction is within $\pm0.1$ mag.}
\label{figmeanRC2}
\end{figure} 
%%%%%%%%%%%%%% figure %%%%%%%%%

To answer this question,
Figure~\ref{figmeanRC2} shows again the population corrections
to the $K$ band for simple stellar populations, but this time it is
also marked the region in which the population correction is within
$\pm0.1$ mag. If a systematic uncertainty of this order on the RC
distance is considered to be acceptable, the allowed range of age
and metallicity is easily evaluated. 
For example, one can obtain distances with this precision, without
evaluating the population correction, for $t$ and $Z$ values typical of the 
sample of Galactic open clusters older than $\sim$2.5 Gyr. 
It is however not allowed to neglect $a priori$ the evaluation of 
$\Delta M_K^{\rm RC}$, due to the fact that 
it can reach values as high as $\pm$0.5 mag.

%%%%%%%%%%%%%%%%%%%%%%%%%%%%%%%%%%%%%%%%%%%%%%%%
\section{$V$-,$I$-, $K$-band population corrections for external galaxies}
\label{sec_galaxies}

Having assessed the adequacy of the theoretical population
corrections for simple stellar populations, 
in this section we 
proceed to compute the corrections 
in $V$, $I$ and $K$ for the composite systems discussed
in Paper~II, using 
the appropriate SFR and AMR discussed in Paper~II; 
the results are summarized in Table~\ref{tab_galaxies}. 
The full references for the SFR and AMR can be found
in table~4 of Paper~II. 

\begin{figure}
\psfig{file=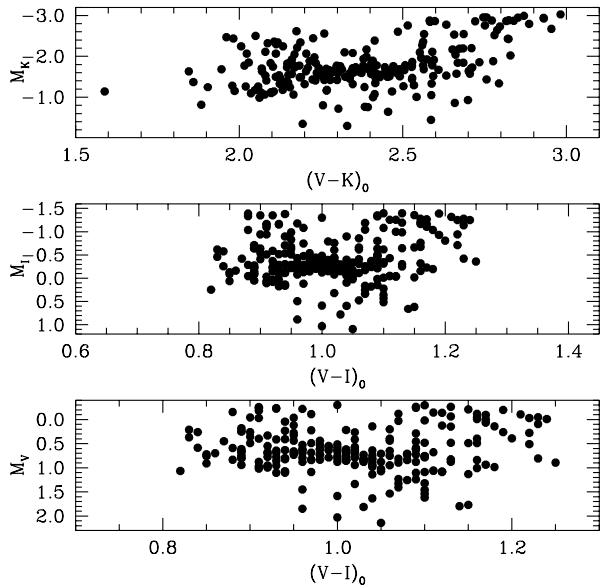,width=8.3cm}
\caption{Empirical CMDs for the local RC (from
Alves~2000) in the $V$, $I$ and $K$ passbands.}
\label{alvesRC}
\end{figure}

\begin{figure}
\psfig{file=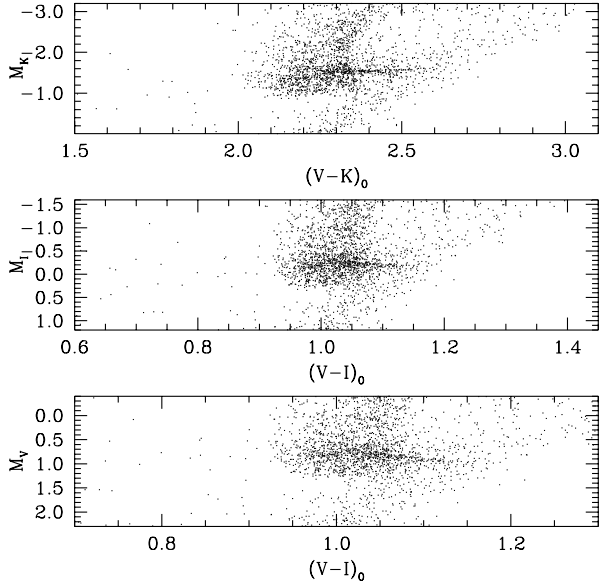,width=8.3cm}
\caption{The same as Fig.~\protect\ref{alvesRC}, but for a
theoretical simulation with a much larger number of stars.}
\label{localRC}
\end{figure}

As a test, we show in Figs.~\ref{alvesRC} and 
\ref{localRC} the empirical data for the local RC (from
Alves~2000) in the $V$, $I$ and $K$ passbands, and the results from
the theoretical simulations, respectively.
Notice the good correspondence between the predicted and observed
shape of the RC in the different colour planes. In particular, the RC
in the $M_I$--$(V-I)$ plane is basically horizontal (see the in-depth
discussion of Paper~II on the local RC and its properties in the 
$M_I$--$(V-I)$ plane), while in the $M_K$--$(V-K)$ plane
is slightly tilted towards higher $M_K$ values, and in
the $M_V$--$(V-I)$ plane is tilted towards lower $M_V$ values.
These properties are well reproduced by the theoretical models.
More in detail, the observed slopes 
$\Delta M_V$/$\Delta (V-I)$ and $\Delta M_K$/$\Delta (V-K)$
are equal to, respectively, 1.46$\pm$0.27 and $-0.63\pm$0.12,
which compare well with the corresponding theoretical values 
1.18$\pm$0.07 and $-0.54\pm$0.02.
In addition, the theoretical simulation predicts
no significant correlations between [Fe/H] and
either $M_K$ or $M_V$, in agreement with the empirical data
by Alves~(2000).

%%%%%%%%%%%%%% figure %%%%%%%%%
%%%%%%%%%%%%%% figure %%%%%%%%%
%%%%%%%%%%%%%% figure %%%%%%%%%
%%%%%%%%%%%%%% figure %%%%%%%%%

The data in Table~\ref{tab_galaxies}, in conjunction with the 
empirical $V$, $I$ and $K$ absolute brightness of the local RC 
reported in Section~\ref{sec_popcorr}, can be used to derive
simultaneously reddening and distances to nearby galaxies, 
following the technique
employed by Alves et al.~(2002). The basic idea is that
the apparent distance moduli determined simultaneously in various  
photometric bands, using the appropriate population corrections, 
must all provide the same unreddened distance; after
assuming a reddening law, one can therefore use this constraint to
derive simultaneously reddening and true distance modulus.
Alves et al.~(2002) have adopted for the reddening law the following
ratios: $A_K$/E(B-V)=0.35, $A_I$/E(B-V)=1.96, $A_V$/E(B-V)=3.24. We
will use these same ratios throughout the rest of this section.

The LMC value of $\Delta M_V^{\rm RC}$ provided in
Table~\ref{tab_galaxies} is slightly different from the one used
by Alves et al.~(2002), which we obtained after a preliminary
evaluation of this quantity.
This small difference (0.04 mag) does not influence appreciably the
results obtained by Alves et al.~(2002). In fact, we repeated their
analysis employing the $\Delta M_V^{\rm RC}$ of Table~\ref{tab_galaxies}
($\Delta M_I^{\rm RC}$ and $\Delta M_K^{\rm RC}$ are unchanged),
and obtained $\mu_0$=18.505$\pm$0.045 and $E(B-V)$=0.079$\pm$0.014, 
almost coincident with  $\mu_0$=18.506$\pm$0.033 and 
$E(B-V)$=0.089$\pm$0.015
obtained by Alves et al.~(2002).

%%%%%%%%%%%%%%%%%%%%%%%%%%%%%%%%%%%%%%%%%%%
\begin{table}
\caption{Population corrections for the clump in
nearby galaxy systems.}
\label{tab_galaxies}
\begin{tabular}{lrrr}
\noalign{\smallskip}\hline\noalign{\smallskip}
System   & $\Delta M_V^{\rm RC}$ &$\Delta M_I^{\rm RC}$ & $\Delta M_K^{\rm RC}$ \\
\noalign{\smallskip}\hline\noalign{\smallskip}
Solar Neighbourhood  & 0.00  & 0.00 & 0.00 \\
({\em Hipparcos}) &  &  & \\
\noalign{\smallskip}\hline\noalign{\smallskip}
Baade's Window & $-0.21$ & $-0.08$ & $-0.07$\\
(scaled-solar) &  & & \\
%%%%%%%%%%%%%%%%%%%% AGGIUNTO %%%%%%%%%%%%%%%%%%
\noalign{\smallskip}\hline\noalign{\smallskip}
Baade's Window & $-$0.06 & $-0.01$ & $-$0.11\\
($\alpha$-enhanced) & & & \\
%%%%%%%%%%%%%%%%%%%% AGGIUNTO %%%%%%%%%%%%%%%%%%
\noalign{\smallskip}\hline\noalign{\smallskip}
Carina dSph & $+0.59$ & $+0.35$ & $-0.17$\\
\noalign{\smallskip}\hline\noalign{\smallskip}
SMC & $+0.31$ & $+0.29$ & $-0.07$\\	
\noalign{\smallskip}\hline\noalign{\smallskip}
LMC & $+0.26$ & $+0.20$ & $-0.03$ \\
%%%%%%%%%%%%%%%%%%%% AGGIUNTO %%%%%%%%%%%%%%%%%%
\noalign{\smallskip}\hline\noalign{\smallskip}
\end{tabular}
\end{table}
%%%%%%%%%%%%%%%%%%%%%%%%%%%%%%%%%%%%%%%%%%%

We have further tested this method 
(and the corresponding population corrections) on the Galactic globular
cluster 47~Tuc, for which an accurate $Hipparcos$-based 
empirical main-sequence fitting distance 
$\mu_0$=13.25$\pm$0.07 has been recently provided by
Percival et al.~(2002), and reddening estimates are also
available ($E(B-V)$=0.04$\pm$0.02, 
see the discussion in Percival et al.~2002).
Spectroscopic analyses give [Fe/H]=$-$0.70$\pm$0.1, together
with an enhancement of the $\alpha$-elements typical of the Galactic
Halo population. We have taken into account the $\alpha$-element
enhancement -- as in Paper~II -- using the results 
by Salasnich et al.~(2000). $V$- and $I$-band data come from the photometry
employed by Percival et al.~(2002), while $K$-band data are taken from
the analysis by Grocholski \& Sarajedini~(2002). We have assumed 
an {\sl a priori} age of 11 $\pm$ 2 Gyr for the cluster, which corresponds 
to the age range allowed by an uncertainty of more than $\pm$0.1 mag
around the main sequence fitting distance. 
By applying Alves et al.~(2002) procedure we have obtained 
$\mu_0$=13.19$\pm$0.07 and $E(B-V)$=0.055$\pm$0.015, in good agreement
with the independent determinations given before.

$K$-band data for a large
variety of Galactic and extragalactic stellar populations
are not yet widely available, while $VI$
data are much more common in the literature. We have
therefore applied Alves et al.~(2002) technique using
only $VI$ data. If, as shown in the previous section, our
population corrections are accurate, one expects to obtain distances 
consistent with the results from $VIK$ data, 
albeit possibly with larger errors
due to the lack of an additional constraint from 
the $K$-band magnitudes.
In case of old metal poor stellar populations,
the RC $V$ and $I$ brightness is much less dependent on age than the
$K$ magnitude, so that the assumptions about the stellar ages are less
critical.
By repeating the determination of 47~Tuc distance with the $VI$ data only,
we obtained $\mu_0$=13.26$\pm$0.07 and $E(B-V)$=0.02$\pm$0.02,
consistent with the result from $VIK$ photometry.
The same exercise can be performed on Alves et al.~(2002) LMC data; we
derived in this case $\mu_0$=18.53$\pm$0.07 and $E(B-V)$=0.071$\pm$0.022,
in very good agreement with the result from the full 
$VIK$ analysis.

As a final exercise, we have studied 
the field LMC-SC6 from the OGLE-II database (Udalski
et al.~2000); this field overlaps with the field
centred on the eclipsing binary HV982, studied by Larsen et al.~(2000).
We have corrected the zero point of the OGLE-II $VI$ photometry, in order
to place it on the same system of Alves et al.~(2002), and employed
stars with random photometric error smaller than 0.04 mag.

We obtain for the RC level in field LMC-SC6 
$V^{\rm RC}$=19.21 and $I^{\rm RC}$=18.17.
By employing the population corrections
given in Table~\ref{tab_galaxies} we have obtained 
$\mu_0$=18.47$\pm$0.04 and $E(B-V)$=0.085$\pm$0.015.
This distance agrees well within the errors with  
Alves et al.~(2002) results.
The minimum reddening for this field as estimated by Larsen et al.~(2000)
is $E(B-V)$=0.085$\pm$0.02. Fitzpatrick et al.~(2002) obtain from spectral fitting
of the eclipsing binary HV982 a value $E(B-V)$=0.086$\pm$0.005.
Oestreicher et al.~(1995) provides a minimum
reddening of 0.05 mag. All these values are consistent with our 
reddening estimate for the RC population.

\section{Conclusions}
\label{sec_conclu}

We have presented a detailed analysis of the behaviour
of the RC brightness in the $K$-band, together with tests
for the accuracy of the predicted population corrections
to the local RC $VIK$ brightness, and a discussion of the Alves et
al.~(2002) method for deriving simultaneously distance and reddening
using multiband RC photometric data.
More in detail:

\noindent

-- we have shown that the $K$-band brightness of the RC is mainly
determined by the behaviour of the bolometric correction to the 
$K$-band. For ages larger than about 1.5 Gyr the trend of the RC
brightness with respect to [M/H] is reversed in comparison with the
$V$- and $I$-band, that is, $M_K^{\rm RC}$ decreases for
increasing metallicity.
By considering the entire [M/H] range spanned by our models 
(Girardi et al.~2000),
for ages below $\sim$4 Gyr $M_K^{\rm RC}$ is on average 
less sensitive to [M/H] than both $M_V^{\rm RC}$ and $M_I^{\rm RC}$. 
At higher ages $M_K^{\rm RC}$ is
more affected by [M/H] than $M_I^{\rm RC}$; for ages larger than
$\sim$11 Gyr it is more metallicity-dependent than also $M_V^{\rm
RC}$, at least for [M/H] lower than $\sim -$0.7.
It is interesting to notice that for [M/H] between 0.0 and
$\sim -$0.4  $M_K^{\rm RC}$ is less affected by the metallicity than 
both $M_I^{\rm RC}$ and $M_V^{\rm RC}$, at least up to ages of about
11 Gyr.
This complex relationship among the RC $V$, $I$ and $K$ brightness
is due to their different sensitivities to age. In particular,
$M_K^{\rm RC}$ is more affected by age than both $M_I^{\rm RC}$
and $M_V^{\rm RC}$, for the lower metallicities and higher ages.
Due to the non-trivial dependence of $M_K^{\rm RC}$ on both [M/H] and age,
it is always necessary -- as in the case of $M_V^{\rm RC}$ and
$M_I^{\rm RC}$ -- to determine the appropriate population
correction for the stellar population under scrutiny, when using 
the RC as a standard candle;

-- we have provided in Table~\ref{tab_mk}
values of the RC average $K$ brightness for a
large range of [M/H] and ages. These data, used in conjunction with 
eq.~\ref{eq_integal}, eq.~\ref{eq_nclgal}, and the data in Table.~1
of Paper~II, can be employed to determine multiband
population corrections for
any given stellar population, once 
SFR and AMR are prescribed;

-- we have simultaneously and positively tested the evolutionary corrections
$\Delta M_V^{\rm RC}$, $\Delta M_I^{\rm RC}$ and 
$\Delta M_K^{\rm RC}$ against Galactic open cluster data;

-- we have discussed the method applied by Alves et al.~(2002) 
for determining reddening and distance from 
the $V$-, $I$- and $K$-band RC brightness; we have positively tested 
this technique on the Galactic globular cluster 47~Tuc, for which both an
empirical parallax-based main sequence fitting distance and reddening
estimates exist. We have also studied the case of using only $V$ and
$I$ photometry, recovering consistent results for both reddening and distance.
Application of this method to an OGLE-II field, and the results by
Alves et al.~(2002), confirm a LMC distance modulus of about 18.50, in
agreement with the $HST$ extragalactic distance scale zero-point
(Freedman et al.~2001).

%%%%%%%%%%%%%%%%%%%%%%%%%%%%%%%%%%%%%%%%%%%%%%%%
\section*{Acknowledgments}
We thank D. Alves for his useful comments on a 
preliminary version of the manuscript.

%%%%%%%%%%%%%%%%%%%%%%%%%%%%%%%%%%%%%%%%%%%%%%%%

\label{lastpage}

\end{document}